\documentclass[RNAAS]{aastex631} 

\usepackage{color}
\usepackage{xspace}

\begin{document}

\title{Satellite Visibility During the April 2024 Total Eclipse}

\author[0000-0001-5368-386X]{Samantha~M. Lawler}
\affiliation{Campion College and the Department of Physics, University of Regina}
\author[0000-0003-1927-731X]{Hanno Rein}
\affiliation{Department of Physical and Environmental Sciences, University of Toronto at Scarborough}
\author[0000-0002-0574-4418]{Aaron C.~Boley}
\affiliation{Department of Physics and Astronomy, University of British Columbia}

\begin{abstract}

On 8 April 2024, tens of millions of people across North America will be able to view a total solar eclipse. Such astronomical events have been important throughout history, but with nearly 10,000 satellites in orbit, we question whether total eclipses will now reveal a sky full of satellites, fundamentally changing this experience for humanity. 
Using the current population of Starlink satellites, we find that the brightest satellites would be naked-eye visible in dark skies, but the high sky brightness during totality will make them undetectable to the unaided eye.
Our model does not take into account the effects of chance reflections from large, flat surfaces like solar panels, which we expect will cause glints and flares that could be visible from large satellites and abandoned rocket bodies. 
Time-lapse all-sky imaging might reveal satellites during the eclipse. 

\end{abstract}


\section{Introduction} \label{sec:intro}

There are currently over 9,500 active satellites in orbit\footnote{See Celestrak at: \url{https://celestrak.org/satcat/boxscore.php}}, of which 59\% are SpaceX's fast-growing Starlink megaconstellation, along with 3,000 inactive satellites, over 2,200 abandoned rocket bodies, and about fifteen thousand trackable space debris pieces.  Moreover, there are an estimated one million pieces of small space debris that are lethal to satellites, but untracked.\footnote{\url{https://www.esa.int/Space_Safety/Space_Debris/Space_debris_by_the_numbers}} 

While all these orbital objects raise serious space safety concerns, they also reflect and scatter sunlight. 
Ground-based light pollution is well-known and a major problem throughout the world. 
But orbital light pollution is fundamentally different. 
It affects the sky globally, changing humanity's connection with the night sky worldwide, without the option of local oversight.
It is also becoming a serious problem for both ground- and space-based astronomy research \citep[e.g.,][]{Mroz2022,Kruk2023}. 

With the upcoming total solar eclipse, tens of millions of people in North American will be able to experience an astronomical event that has inspired awe throughout human history. In addition to eclipses facilitating tests of scientific theories \citep[e.g.][]{Dyson1920},
estimates place the economic potential to be 1 billion USD from tourism\footnote{\url{https://www.forbes.com/sites/jamiecartereurope/2024/02/24/us-could-get-1-billion-boost-from-april-8s-total-eclipse-of-the-sun/}}, similar to the economic impact of the Super Bowl.\footnote{\url{https://www.businessinsider.com/economics-super-bowl-las-vegas-billion-dollars-new-orleans-2024-2}} 
Eclipses continue to be culturally significant events, throughout the world.

With so much interest, we ask whether satellites are expected to change this shared experience by creating a multitude of additional lights moving across the sky during totality. 
Trying to answer the question has revealed some interesting complications: we expect satellites to be dim and undetected, although there remain situations for which some objects could be visible.

\section{Brightness simulation} \label{sec:sim}

We approach the problem using similar methods to those outlined by \citet{Lawler2022}. 
Normally when making satellite brightness predictions, we ask whether the space object is in Earth's shadow, and if not, we estimate its brightness using its range, albedo, surface area, and phase (the angle between the observer, the satellite, and the Sun). 
The actual reflectance also depends on the satellite's attitude, along with its bidirectional reflectance distribution function. 
We typically do not have such this information, and so we prefer a simple phase model.
Specifically, we use a Lambertian sphere \citep[see][]{Lawler2022}, informed by our Starlink observations at night \citep{Boley2022}. We also tested several other prescriptions that do not change our overall conclusions. 

Daytime observation during an eclipse requires two important modifications to our method:
First, the satellites that are closest to the observer and would normally be the brightest, are also in the Moon's shadow, so only a fraction of the incoming sunlight is available to be reflected.
Second, while the satellites are partly shielded from the Sun, they also receive Earthshine, and so even satellites that are entirely within the Moon's umbra will be partially illuminated. 

For our models, we use Starlink TLEs obtained from CelesTrak\footnote{\url{https://celestrak.org}} on 18 March 2024. 
Those TLEs are then propagated to the time of mid-totality for an observer in Kingston, Ontario.
We recognize that these TLEs are not suitable for propagation so far into the future, but the results are sufficient in a statistical sense. 

\section{Sky predictions} \label{sec:pred}

\begin{figure} \centering
\includegraphics[height=0.6\textwidth]{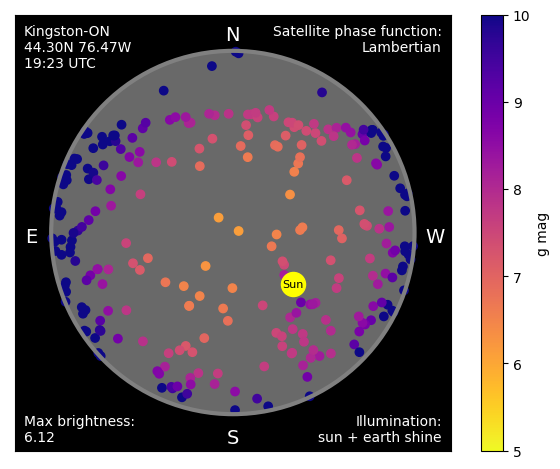}
\caption{All-sky visibility plot for current Starlink satellites during totality, observed from Kingston, Ontario. Color shows visible $g$-magnitudes. Overall, satellites get gradually brighter toward zenith as the range decreases, with the shadow of the Moon causing fainter satellites close to the position of the Sun in the sky.  
\label{fig:sky}}
\end{figure}

Figure~\ref{fig:sky} shows a predicted snapshot of the satellite brightness distribution for Kingston, Ontario during totality.  
The reflected light seen by the observer from each satellite is reduced by the fraction of the Sun that is covered by the Moon, as seen from the satellite's position, with additional satellite illumination coming from the Earth below.
Because parts of Earth are also in the Moon's shadow, we model the Earthshine effect using a ray tracing algorithm that takes into account the complex illumination geometry.

We find that illuminated satellites during the eclipse should be faint due to the combination of very large solar phase angles and satellites in the sky falling inside the Moon's penumbra.
The brightest satellites in our simulated sky are predicted to have apparent $g$-magnitudes near 6.1.  
This result is much fainter than our predictions at night/twilight, which have smaller phase angles and no Moon-shadowing, resulting in predicted $g_{\rm max}$ as bright as 3.5 \citep{Lawler2022}.  

$g$=6.2 would be naked-eye visible in a dark night sky, but the sky brightness during totality will be brighter than on a full-moon night, 
12.5 mag/arcsec$^2$ \citep{Pramudya2016}. 
Thus we do not expect any Starlink satellites to be naked-eye visible during the eclipse.

However, if Starlink satellites are in a ``shark-fin'' orientation (solar panels perpendicular to the Earth's surface), there could be significant short-lived reflections (``glints'') off the large, flat solar panels.  
These reflections have already been observed in research images \citep{Karpov2023}.  
Because satellite information concerning orientation, shape, or materials properties is not generally shared, we are unable to predict glints at this time.
Glints will vary for each observer due to the brief geometric alignments involved.  

Although this simulation focuses on Starlink only, since they comprise the vast majority of active satellites, there are thousands abandoned rocket bodies currently in orbit, as noted above.
These large objects stay in orbit for days to many decades after launching their payload.
They can be very bright while at large distances from the observer, and thus can be visible throughout the night and year \citep[see][chpt.~4]{byersboley2023}. 
They also tumble chaotically, producing bright glints that can be visible to the naked eye.\footnote{The authors have an ongoing program assessing rocket body brightness and variability.}
We suspect that such objects will be visible during an eclipse, but require observations for verification. 

\section{Conclusion}

Bright satellites are changing the sky worldwide, disrupting astronomy research and cultural stargazing alike.  Our simulations here show that sunlit Starlink satellites will not be naked-eye visible in the darkened sky during the April 2024 total eclipse.  Glints from artificial objects in orbit could still be visible. 
Wide field-of-view time-lapse observations during totality may prove valuable in determining the magnitude and frequency of these glints, as well as provide further information for assessing satellite brightness.
The sky is changing rapidly due to the commercialization of LEO, and we invite astronomers reading this to join groups advocating for regulation of orbit, like IAU-CPS\footnote{\url{https://cps.iau.org/}}.

\software{This research was made possible by the open-source projects \texttt{Skyfield} \citep{2019ascl.soft07024R}, \texttt{REBOUND} \citep{ReinLiu2012}, \texttt{Jupyter} \citep{jupyter}, \texttt{iPython} \citep{ipython}, and \texttt{matplotlib} \citep{matplotlib}.}

\vspace{5mm}

\newpage
\bibliographystyle{aasjournal}


\end{document}